# Holonomic control of a three-qubits system in an NV center using a near-term quantum computer

Shaman Bhattacharyya and Somnath Bhattacharyya[a)]

Nano-Scale Transport Physics Laboratory, School of Physics, University of the Witwatersrand, Johannesburg, 1 Jan Smuts Avenue, Wits 2050, South Africa

The holonomic approach to controlling (nitrogen-vacancy) NV-center qubits provides an elegant way of theoretically devising universal quantum gates that operate on qubits via calculable microwave pulses. There is, however, a lack of simulated results from the theory of holonomic control of quantum registers with more than two qubits describing the transition between the dark states. In light of this, we have been experimenting with the IBM Quantum Experience technology to determine the capabilities of simulating holonomic control of NV-centers for three qubits describing an eight-level system that produces a non-Abelian geometric phase. The tunability of the geometric phase via the detuning frequency is demonstrated through the high fidelity (~80%) of 3-qubit off-resonant holonomic gates over the on-resonant ones. The transition between the dark states shows the alignment of the gate's dark state with the qubit's initial state hence decoherence of the multi-qubit system is well-controlled through a π/3 rotation. The electron return probability can exhibit spin-orbit coupling-like behavior as observed in topological materials based on the extra geometric phase.

The supposition that non-Abelian geometric phases can be applied to holonomic control of qubits was postulated over two decades ago [1]. Since then, holonomic control of qubits has progressed from speculation to the realization of control of physical qubits using their non-Abelian geometric phases [2-9]. The idea behind this rotation of qubits is the creation of a non-Abelian geometric phase. Such a geometric development can provide a robust way toward universal quantum computation since the geometric phases are determined by the global properties of the evolution paths and possess a built-in noise-resilience feature against certain types of local noises. Such geometric phases have been implemented by a single cycle of non-adiabatic evolution which is achieved by rotating a single qubit with a holonomic gate [10-12]. Further progression is seen in the development of multi-qubit holonomic gates, both in superconducting qubits and ion traps [6-8]. Nitrogen vacancy (NV) centers in diamonds show great promise for holonomic control due to having both a stable non-Abelian geometric phase and the means of control via optical excitation [10-14]. Qubits in NV centers can be rotated by microwave, radiofrequency, and Stokes laser pulses [13-18]. These pulses can be used to implement a set of universal quantum gates which then operate on the qubits. Such control of NV centers via non-Abelian geometric phases has been demonstrated, with holonomic single-qubit gates implemented on NV centers demonstrating high fidelity, even at room temperature [11,12].

Recently, the use of polarised microwaves on NV centers has demonstrated the ability to perform universal quantum gate-based operations on multiple qubits [9]. Through the creation of two separate cycles on two separate qubits which are then concatenated a geometric phase will be created [19,20]. However, there are very few attempts to show the holonomic control of three coupled qubits [21,22,23,24]. In particular, a method of performing holonomic control of three-qubit systems (three three-level Rydberg atoms) using a single holonomic gate is proposed however, NV centers have specific energy levels of transition [23]. In our previous report, we demonstrated a quantum simulation of two levels of an NV center by an IBM QE [25]. Recently, we also showed three qubits connected by three coupling parameters without considering any geometric phase [25]. In this work, we demonstrate the holonomic control of three coupled qubits in addition to the replication of some well-known results which provides a solid foundation for simulating holonomic control of NV-centers using IBM QE [see Ref. 10,12, and 22 and Supplementary information].

A three-qubit system produces a similar effect to Rashba spin-orbit coupling (RSOC) and vortex structures in materials which are highlighted in the paper [26,27,28]. There have been attempts to acquire the geometric phase in a qubit circuit of IBM QE [29,30,31]. Also, the improved return probability ($P_r$) of electrons in the asymmetric arrangements of three entangled qubits has been demonstrated [28]. However, a set of three qubits accumulating a geometric phase dependent on time has not been achieved [32-35]. A holonomic gate can be defined as an operation or gate which causes a qubit rotation such that the qubit acquires an Abelian phase which can be implemented on qubits in optical systems such as NV centers or superconductors [36-40]. For the three-qubit system in a trigonal arrangement that occupies a closed space, we can generate a non-Abelian geometric phase and show the superiority of off-resonant holonomic gates over the on-resonant gates. This development shows particularly interesting potential as NV centers are viable qubits at room temperature, and the operation of universal gates on such qubits is a step forward in the development of universal quantum computers.

Another very important property of a holonomic gate in an NV center is that it also has a dark and bright state. This property and the initialized state of the qubit play a role in the gate's performance by controlling the level of decoherence in the system. According to Ref. 10, the decoherence for a single qubit system reaches a minimum when the dark states of the gate align with the initial state of the qubit. For 2 qubits or a 4-level system $|D_1\rangle$ and $|D_2\rangle$ were demonstrated [18]. Here, we show the transition between three dark states in a three-qubit system.

[a)] Electronic mail: somnath.bhattacharyya@wits.ac.za

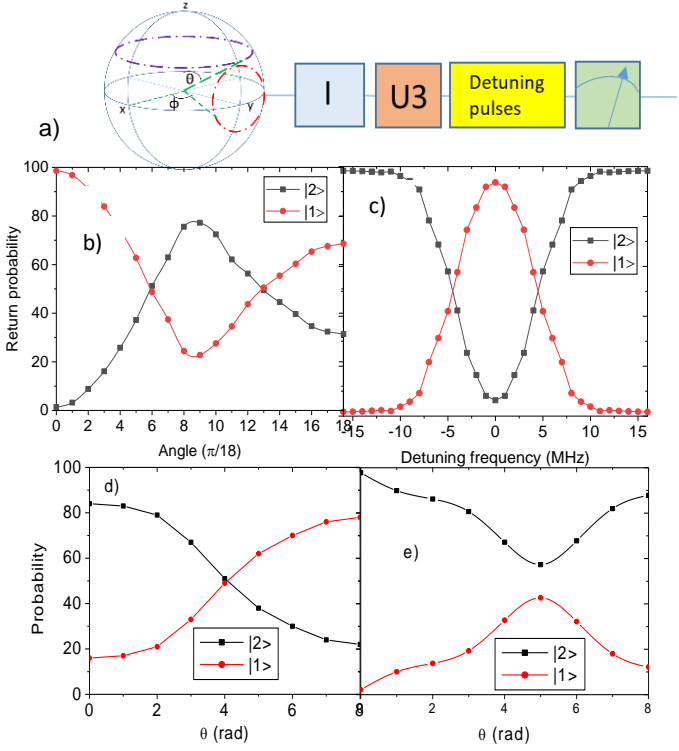

**Fig 1**: Simulation of holonomic control on a single-qubit. **a)** The rotation path is implemented by a single holonomic gate. Since the path forms a closed-loop a geometric phase can be created via a single iteration. The circuit shown here is used to initialize the qubit and then implement the holonomic gate using a pulse schedule. Then the system is measured on the $|z\rangle$ basis. **b)** Gates with the angle $\theta$ variable after initializing the qubit to $|1\rangle$. The return probabilities of the final states are to be measured on the $|z\rangle$ basis from both $|1\rangle$ and $|2\rangle$ and to be plotted against $\theta$. The red and black lines indicate the behavior of the qubit in the presence of noise. **c)** A holonomic gate with a variable detuning frequency and fixed rotation axis x is applied to the qubit. The oscillations for the qubit are shown which are caused when the detuning frequency causes the qubit to rotate around the x-axis. The red and black lines indicate emulations in the presence of decoherence. The phase change becomes more observable when the qubit is initialized to the state $|1\rangle$. **d)** Emulation of multiple single qubit holonomic gates. Gates with the angle $\theta$ and $\phi$ variable where $\theta = \varphi$. The qubit is initialized to 0 and the rotation axis is kept constant since $\theta = \varphi$. The probability of the final states is obtained using the $|z\rangle$ basis for measurement. For the implementation of the rotation, we use an $R_X$ and an $R_Z$ gate. **e)** The same set of holonomic gates is implemented on the qubit, however, the qubit is initialized by applying a $\pi/4$ rotation to the state $|1\rangle$. Red and black lines indicate emulations with decoherence. Both the phase and decoherence are reduced when the rotation axis of the qubit aligns with the dark state of the gates since it undergoes trivial dynamics.

To implement holonomic control on multiple qubits, we first demonstrate the operation of a holonomic gate on a single qubit. The circuit for the implementation of a **one-qubit** holonomic gate consists of three main processes, namely the initialization of the qubit, the holonomic gate implementation, and the measurement [Fig. 1a)]. We start by initializing a Bloch sphere to the states $|z\rangle$ and $|-z\rangle$ since the final states of the qubit is measured on the standard $|z\rangle$ basis. We also initialize the standard states $|x\rangle$ and $|y\rangle$. Unlike an ordinary qubit where $|z\rangle = 0$ and $|-z\rangle = 1$, in this Bloch sphere the state $|z\rangle = |1\rangle$ and the state $|-z\rangle = |2\rangle$. A set of pulses is used to implement a single qubit unitary holonomic gate $U$ with parameters $\theta$, $\varphi$, and $\lambda$. Here $\theta$ and $\varphi$ are used to denote the rotation angles in the Bloch sphere and $\lambda = \Delta/\Omega$ where $\Delta$ and $\Omega$ correspond to the detuning frequency of the pulse and Rabi frequency of the qubit, respectively. The holonomic gate on IBM QE $\theta$ and $\varphi$ is implemented using $R_X$, $R_Y$, and $R_Z$ gates, respectively. The pulse with the frequency $\Delta$ is appended into the circuit to achieve the detuning. To this end, we reproduce some of the previous work through emulations on IBM QE [10, 11].

Now, we continue by characterizing arbitrary rotations around the x-axis via an $R_X$ gate and other gates of the set $X(\gamma)$ with $\varphi = 0$ where the phase shift $\gamma$ changes as a function of the detuning frequency [Fig. 1b)]. This includes the X gate which corresponds to $\gamma = \pi$. This is extracted from the difference between the input and output state and expanded to show the holonomic rotations around arbitrary axes. We keep $\varphi$ constant while $\theta$ is the variable. The discrepancy between the emulated and simulated transfer increases with $\theta$ starting from 0 and attains the largest values when $\theta = \pi$ as the initialized state aligns with the dark state of the qubit. It can be concluded that the difference between the emulated and simulated result increases as $\theta$ increases but the discrepancy can be reduced if the initial state aligns with the dark state of the holonomic gate. The rotation of the qubit with variable detuning frequency is shown in Fig. 1b) and c) by measuring the $P_r$ in the $|1\rangle$ and $|2\rangle$ basis, with the most observable phase change seen when the qubit is initialized to $|1\rangle$.

By changing the axis of rotation and by increasing angle $\theta$, keeping $\varphi = 0$ constant, we demonstrate effective Rabi oscillations between $|1\rangle$ and $|2\rangle$ as shown in Fig. 1b). Similarly, by keeping $\theta = 0$ constant and changing $\varphi$ Rabi oscillations can be demonstrated [15], implying holonomic control over the $(\theta, \varphi)$ degrees of freedom [Fig. 1b) or c)]. We also demonstrate tuneable rotations of an NV center around the x-axis and y-axis using a variable detuning frequency ($\Delta$) that produces a phase shift in the qubit. The detuning frequency is varied over a range of frequencies that span resonance, resulting in the Rabi oscillations shown in Fig. 1c). Since the holonomic gate is represented by a unitary matrix which is part of the SU(2) group the actual phase accumulated cannot be directly measured. The discrepancy between the emulated and simulated results can be used to calculate the accumulated phase. However, when we keep detuning the frequency of a variable and hold $\theta$ at $\pi/2$ we realize that the discrepancy is the greatest at $\Delta = 0$ and also varies as a function of $\Delta$. The measured fidelities of the $X(\pi/2)$ and $Y(\pi/2)$ are 0.84 and 0.86, respectively.

To determine the potential performance of the single-qubit holonomic gate in IBM QE we also use sequences of three gates to achieve holonomic rotation as shown in Fig. 1d) and e). Using the same techniques for holonomic control already demonstrated, we emulate two further single-qubit rotations in the $|z\rangle$ basis. The $P_r$ are shown in Fig. 1d) and 1e). The first rotation intersects the states $|1\rangle$ and the $1/\sqrt{2}(|1\rangle + i|2\rangle)$ on the surface of the Bloch sphere. The second rotation traverses the circumference of the Bloch sphere about an axis defined by an angle $\theta = \pi/4$. After initializing to state $|1\rangle$, it can be seen that the gate transfers the population from $|1\rangle$ to $|2\rangle$ around the x-axis and back to $|1\rangle$ as the frequency of the pulse is altered. For the composite holonomic control, we observe that the maximum discrepancy between the emulated and simulated results is lower than that of the single holonomic gate control. This shows that the accumulated phase from multiple rotation gates is smaller than that of the single holonomic gate. Also, the fidelities of the composite gates reach an average of 0.77 which is lower than the single gate. It can also be concluded that using a single loop scheme is generally more efficient than using a complex gate that concatenates separate cycles because the fidelity of the single loop scheme (of around 0.87) is higher than the fidelity of 0.66 of complex gates. However, the use of multiple holonomic gates on one qubit has a similar efficiency compared to a single unitary gate. This analysis supports the results shown previously [10].

For **two qubits** the holonomic control is demonstrated in a four-level system [9]. We have considered oscillations between two ground states ($|1\rangle = |00\rangle$ and $|2\rangle = |01\rangle$) and two excited states ($|3\rangle = |10\rangle$ and $|4\rangle = |11\rangle$) in a negatively charged NV-center. Using these approximations as the parameters of an $R_x$ gate on each qubit of a two-qubit circuit results in the required $\pi/2$ gate being performed on the qubits Fig. 2a). The holonomic control of a 4-level system has been demonstrated in Fig. 2b). When attempting to simulate this on IBM QE, the envelope and wavefunction had to be programmed (see Figs. S1 and inset in Supplementary information) [12].

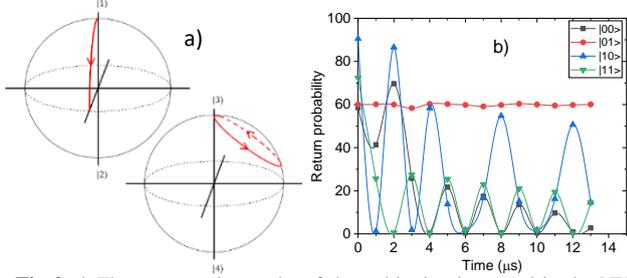

**Fig 2 a)** The two rotation paths of the qubits implemented in the NV center. These paths start and end at the same position on each qubit. The first path is implemented on the first qubit while the second path is implemented on the second qubit. **b)** The return probability of the state |1> after evolution in the rotation path for varying degrees of the degeneracy of the ground states over time.

In the **three-qubit** holonomic control procedure, we use the GHZ frame of reference when mentioning any qubit states. Now in reality it is impossible to demonstrate all the nine levels of an NV center using qubits because for N number of qubits the total number of levels is $2^N$. To get around this problem most holonomic simulations omit the |0> state and replace the poles of the first qubit with the states |1> and |2> (this corresponds with most holonomic simulations). These states can be degenerate. The system for the three-qubit system in the rotating frame is given as:

$$H = \sum_{i=1}^{8}|\omega_i|i\rangle\langle i| + \frac{i}{2}(\Omega_{P_1}e^{-iV_{P_1}t} - \Omega_{P_2}e^{-iV_{P_2}t})(|1\rangle\langle 7| - |1\rangle\langle 8|) - \frac{i}{2}(\Omega_{S_1}e^{-iV_{S_1}t} - \Omega_{S_2}e^{-iV_{S_2}t})(|2\rangle\langle 7| + |2\rangle\langle 8|) + \frac{i}{2}(\Omega_{P_3}e^{-iV_{P_3}t})(|1\rangle\langle 3|) - \frac{i}{2}(\Omega_{S_3}e^{-iV_{S_3}t})(|2\rangle\langle 4|) \quad (1)$$

The Hamiltonian in the interaction picture is given by:

$$H_I = \begin{pmatrix} 0 & 0 & i\Omega_1/2 & 0 & 0 & 0 & i\Omega_2/2 & i\Omega_3/2 \\ 0 & 0 & 0 & 0 & 0 & -i\Omega_6/2 & -i\Omega_5/2 & -i\Omega_4/2 \\ -i\Omega_1^*/2 & 0 & \Delta_1 & 0 & 0 & \Delta_1 & 0 & 0 \\ 0 & 0 & 0 & 0 & 0 & 0 & 0 & 0 \\ 0 & 0 & 0 & 0 & 0 & 0 & 0 & 0 \\ 0 & i\Omega_6^*/2 & \Delta_1 & 0 & 0 & \Delta_1 & 0 & 0 \\ -i\Omega_2^*/2 & i\Omega_5^*/2 & 0 & 0 & 0 & 0 & \Delta_2 & 0 \\ -i\Omega_3^*/2 & i\Omega_4^*/2 & 0 & 0 & 0 & 0 & 0 & \Delta_3 \end{pmatrix} \quad (2)$$

Here $\Delta_1$, $\Delta_2$, and $\Delta_3$ are the detuning frequencies and $\Omega$ is maintained at 15 MHz. The coupling between the ground states and the $E_x$, $E_y$ states (in Fig. 4a)) was not considered because such a coupling would require that all the nine states of the NV center be used which is not possible given the fact that 3 qubits can only encode a maximum of 8 states [|1>=|000>, |2>=|001>, |3>=|010>, |4>=|011>, |5>=|100>, |6>=|101>, |7>=|110>, |8>=|111>.]. We set different values for $\Delta_i$ and solve the Schrödinger equation with the system in the $m_s = 1$ states. In terms of eigenstates, the dark states can be defined as the eigenstates with an eigenvalue of zero.

$$|D'\rangle = (\sin\beta\, e^{-i\varphi}|1\rangle + \cos\beta\, e^{i\varphi}|2\rangle)$$
$$|D''\rangle = (\sin\beta\, e^{-i\varphi}|3\rangle + \cos\beta\, e^{i\varphi}|4\rangle)$$
$$|D\rangle = (|D'\rangle + |D''\rangle)/\sqrt{2} \quad (3)$$

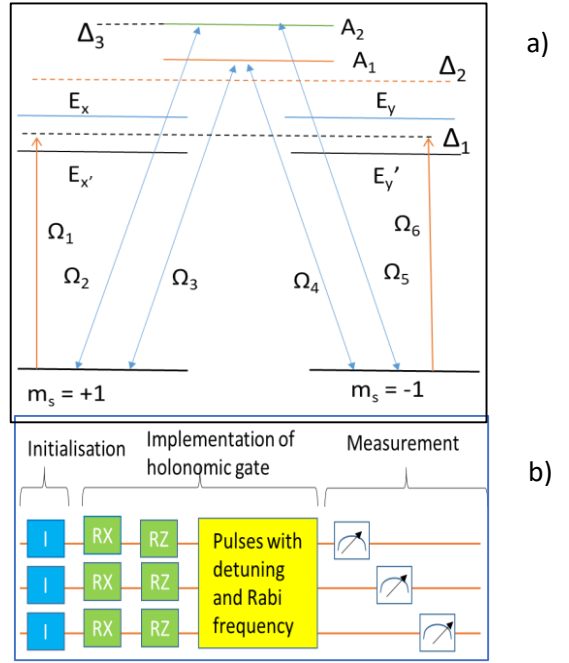

**Fig. 3 a)** Energy diagram of the spin triplets $^3A$ and $^3E$ for the complete 9 level NV center including the selection rule. The Rabi frequencies are $\Omega_{pi}$ for the applied pump and $\Omega_{si}$ for the Stokes field with $i = 1, 2,$ and $3$. The detuning is $\Delta_1 = \Delta_2 = \Delta_3 = \Delta$. **b)** Quantum circuit used for the holonomic control of three qubits. The identity gates (any other gate that can be used here) are used to initialize the qubits. $R_X$ and $R_Z$ gates are used to set the parameters $\theta$ and $\varphi$, respectively. The pulses are implemented with $\Delta_i$ and $\Omega$ on all three qubits to achieve full holonomic control.

In our simulation, we focus on the representation using the states |1> and |2> since it corresponds with the dark and bright states of an NV center. Since there are three coupled qubits and each has a dark and bright state we can represent the triplet nature of the ground states (or nuclear spins) of the NV center. By encoding the nuclear spins in one qubit, the electron spin states in the second and intermediate states into the third qubit we propose a method for holonomic control of three qubits. For the holonomic control, we choose two different scenarios that we simulate. In the same manner, the qubits are initialized to the states |z> and |–z>. The qubit with the states |1> and |2> can be rotated around an arbitrary axis $n$ given as $n = \sin 3\theta\, \cos 3\phi, \sin 3\theta\, \sin 3\phi, \cos 3\theta$. All holonomic gates used in this simulation are unitary operations defined in the same way as for the one qubit holonomic control $U_C = e^{\frac{i\gamma}{2}}|-D\rangle\langle -D| + |D\rangle\langle D|$ (ket $|-D\rangle$ is orthogonal to the ket $|D\rangle$). Hence using this operator any arbitrary rotation of the qubit around any axis is possible. To establish $\pi/2$ rotations around the standard x-axis we

determine the fixed angle $\theta = \pi/6$ and $\varphi = 0$. The rotation around the y axis for $\pi/2$ can also be similarly obtained by setting $\theta = \pi/6$ and $\varphi = \pi/6$.

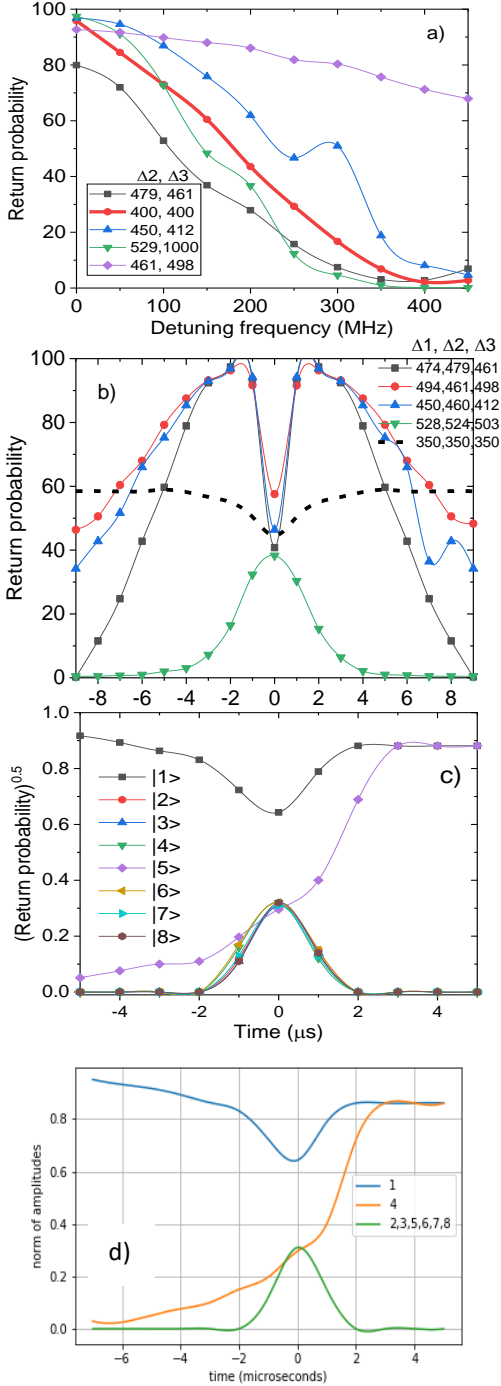

**Fig 4 a)** Simulation of holonomic control on an eight-level NV center with detuning frequency variable for varying detuning frequencies of the three qubits. The rotation paths of each qubit are concatenated to form a single effective rotation path. The probabilities of the state |1> are plotted against detuning frequency $\Delta_1$ for fixed detuning $\Delta_2$ and $\Delta_3$. The solid line with open symbols represents the on-resonant case. **b)** Evolution of the state |1> over pulse sequence duration for varying values of detuning frequencies $\Delta_1$, $\Delta_2$, and $\Delta_3$. The solid dashed line represents the on-resonant case. **c)** The time evolution of the states over pulse duration illustrating a $\pi/3$ rotation. The grey square, purple diamond, and green triangles and lines represent the |1>, |5>, and the |2> states, respectively. **d)** Evolution of the state |1> as a function of time with varying values of detuning frequency as described by the interaction Hamiltonian in equations 1 and 2.

In the first scenario [Fig. 4a)] we only show the holonomic control of the qubits around the standard x, y, and z axes. We use an $R_X$ gate to rotate the first qubit by $\theta = 5\pi/6$. We do the same for the second qubit however, the parameter for this $R_X$ gate is $-\pi/6$. The third qubit is rotated by $\pi/3$ ($\theta = 2\pi/3$). These steps above are used to initialize the system. In a similar manner to the two-qubit scenario above, we consider three rotation paths. For the first qubit, we implement a holonomic gate with $\varphi = 0$ and $\theta = 2\pi$ which is a complete rotation around the y-axis. The Rabi frequency for the first qubit $\Omega_1$ is equal to the qubit frequency stated in IBM QE at 4966 MHz. A similar holonomic gate was implemented on the second qubit. For this holonomic gate $\theta = -2\pi$ which is a complete rotation in the opposite direction from the first qubit. The third qubit is rotated around a circular path around the y-axis with $\theta$ ranging between $\pi/6$ and $-\pi/6$. The same is true for $\varphi$. We keep the detuning frequency for the first holonomic gate $\Delta_1$ variable for varying values of the detuning frequencies for the other holonomic gates $\Delta_2$ and $\Delta_3$.

The paths are concatenated and the system is measured on the z basis to obtain the $P_r$ of the state |1>. In Fig. 4a) a simulation shown by the solid red line represents the ideal scenario where $\Delta_1 = \Delta_2 = \Delta_3$. The return probabilities of the |1> state start at 100 and decrease to 0 as the rotation angle is $\pi$. We also notice that there is a discrepancy between the emulated and simulated results. As mentioned for the one (and two-qubit) scenario, this discrepancy indicates that an effective phase is accumulated across all three paths. The nature of this phase is Abelian for each path. However, the combined effect of all the paths produces a non-Abelian geometric phase because both d$\theta$ and d$\varphi$ are non-zero. The magnitude of the phase varies as a function of the detuning frequency. The magnitude reaches a maximum value at $\Delta_1 = 300$ MHz. The phase increases with the difference between the values of $\Delta_2$ and $\Delta_3$. This is consistent with the fact that off-resonant holonomic gates have a higher performance compared to on-resonant gates which do not allow significant phase accumulation.

In the second scenario [Fig. 4b)] we consider a similar configuration for the system however, we measure the effective rotation time for the varying values of $\Delta_1$, $\Delta_2$, and $\Delta_3$. We measure the rotation time of each of these paths as well as the effective rotation path to obtain the $P_r$ of the state |1>. Each rotation accumulates a geometric phase which is Abelian so that d$\varphi = 0$ for the first two rotation paths, and only either d$\varphi$ or d$\theta$ is non-zero for the whole rotation. Now, since calculating the exact phase matrix requires the evaluation of path integrals and the calculation of gauge potentials, we approximate the effective phase matrix for each azimuthal rotational path of the 1st and 3rd qubit to be:

$$U_{eff} = \begin{pmatrix} 0{,}99 + 0{.}47i & -0{,}82 + 0{,}12i \\ 0{,}93 + 0{,}82i & 0{,}65 + 0{,}33i \end{pmatrix} \quad (4)$$

Although each separate rotation represents Abelian geometric phases, the effective phase can, however, be realized non-Abelian. Our results are consistent with Ref. 22.

In qubits, a dark state is a state where a qubit undergoes trivial dynamics (i.e. the qubit can be described by the qubit Hamiltonian). For the creation of geometric phases, the rotational path traversed by the qubit is usually a closed loop and can be implemented on an arbitrary axis such as the one shown in Fig 1a). The evolution of dark states for a two-qubit system is given in the supplementary information {see Fig.

S1}. In Fig. 4a) it can be observed that the initial return probability at zero $\Delta_1$ for the dark states is the same for all values of $\Delta_2$ and $\Delta_3$. Also, the deviation from the classical approximation is the least at zero $\Delta_1$. This indicates that the dark state of the gate is aligned with the initial states of the qubits. This is consistent with the results obtained for a single qubit in Ref. 10. It is interesting to note that with the increase of the detuning frequency more deviations occur in the return probability. Under normal circumstances, this deviation is mostly related to decoherence in the systems. However, due to the nature of the qubit rotation, this deviation is also related to the creation of a geometric phase. This geometric phase is observed to be partially dependent on the detuning frequency. Also, the gate's dark state seems to align with the initial state of the qubit at 450 MHz as the deviations decrease to a minimum.

In Fig. 4b) a similar conclusion can be drawn regarding the alignment of the dark state of the holonomic gate and the initial state of the qubits. At $t = 0$ the dark state of the gate aligns with the initial state of the qubit. This can be seen by the fact that at $t = 0$ the return probability is the same as the $P_r$ of the simulated results. Both Fig. 4a) and 4b) indicate that the alignment of the dark and initial states allows for the control of decoherence in a system consisting of multi-qubits. Fig. 4c) shows the evolution of the states of the system caused by rotating the second qubit in the system. Only one qubit is rotated so that this rotation can be described by the quantum Rabi model and subsequently a Rabi envelope can be developed. It can be seen that the amplitude of the state |5> increases to 0.86 while the state |1> drops to 0.86. This shows that a $\pi/3$ rotation has been achieved. The excited states do not have a non-zero population once the $\pi/3$ rotation is complete. Also, it is seen that the other states follow the same trend. The increase in amplitude for those states is consistent with the transition between the three dark states of our qubit system. To prove the presence of a dark state in a system the eigenstates of the interaction Hamiltonian using Eq. 2 are calculated as shown in Fig. 4d) which is consistent with Fig. 4c). The 3-qubit off-resonant holonomic gates show higher fidelity (~80%) compared to the on-resonant gates (~70% fidelity).

Dark states can be created by the strong overlapping of two orthogonal states like the spin-orbit interaction. In condensed matter physics it is well known that the geometric phase arises from the RSOC. This, however, has not been experimentally demonstrated using a quantum simulator that requires at least three qubits. Earlier, a three-qubit system was employed to generate a synthetic magnetic field, a chiral ground state current, or a chiral spin state however, the geometric phase was not captured which was attempted in the present work [29]. The purpose of the holonomic operation is to create a geometric phase through a closed-loop through multiple iterations. Dark states can be described as the bound states or the vortex core which arises from the strong spin-orbit coupling which results in weak (anti-) localization phenomena without breaking the time-reversal symmetry as observed in Fig. 4b) [31]. This picture becomes very interesting for three vortices that can be controlled by a single operation. We see fine control of the $P_r$ with the Rabi frequency (or RSOC strength). At the zero frequency, $P_r$ has a peak at the origin or zero detune time (similar to the weak localization phenomena observed in the magnetic field-dependent resistance of metal with the effect of RSOC). This becomes a minimum by the application of the Rabi frequency which is similar to weak anti-localization [31]. This is a hallmark feature of RSOC in a topologically protected material. While two holonomic operations perform a weak localization-like feature, the third holonomic gate breaks the time-reversal symmetry. Hence the operation of a 3-qubit system works more efficiently than the 2-qubit system [12].

The one qubit system can only be encoded with the ground states of the NV centers (or with one dark and one bright state). This means that only one dark state can be measured. However, since the ground states are degenerate no transitions are possible. For the 3-qubit system, since there are more dark states, transitions between the dark states can be observed just through the holonomic control of one of the 3 qubits in the system. The transition is also quite interesting because it causes the amplitudes of the states other than the initial states to increase for a short duration. Also with 3-qubit systems, the system can exhibit geometric dependence which is not possible for one and two-qubit systems since only one configuration exists for these systems while there are two configurations for the qubits in the three-qubit system.

IBM Quantum Experience is a powerful tool for simulating quantum mechanical systems such as NV-centers in diamonds, and the potential for simulating holonomic control of diamond-based spin qubits is very real. Before attempting the holonomic simulations of non-Abelian gates, the basic method of simulating NV centers has been demonstrated. Subsequently, the results of several different papers were reproduced using IBM QE simulations. We use an eight-level system and give a nearly complete description of an NV center. This can be compared to two four-level systems each corresponding to a vortex (a bound state) and their interactions. A set of three qubits also includes strong spin-orbit-like interactions and we see all three dark states of the holonomic gates can be aligned with the initial state of the qubits hence allowing control of the decoherence of the entire system through a $\pi/3$ rotation. For multiple qubits, the Dicke model is required which is beyond the scope of our work. This simulation can also be applied to understand the resonant properties of a three-quantum dot system [41]. Since this is the first attempt at simulating the three-qubit holonomic gate, details of the calculation of the exact nature of this system will be given elsewhere.

## SUPPLEMENTARY MATERIAL

The Supplementary Material includes the methodology and plots of return probability as a function of time for a two-qubit system.

## ACKNOWLEDGMENTS

The authors acknowledge the initial work of D. Mahony, R.M. Erasmus, and N. Bronn (IBM) for discussion. We are very thankful to F. Nori and Y. H. Chen for very carefully reviewing the manuscript. The research grants from the NRF, South Africa under the RSA-India bilateral program and the CSIR-NLC rental pool project are acknowledged.

## CONFLICT OF INTEREST

The authors have no conflicts to disclose.

## DATA AVAILABILITY

The data that support the findings of this study are available from the corresponding authors upon reasonable request.

**References:**

[1] P. Zanardi and M. Rasetti, *Phys. Lett. A* **264**, 94 (1999).
[2] M. A. Kowarsky, L. C. L. Hollenberg, and A. M. Martin, *Phys. Rev. A* **90**, 042116 (2014).
[3] J. Zhang, S. J. Devitt, J. Q. You, and F. Nori, *Phys. Rev. A* **97**, 022335 (2018); J. Zhang, S. J. Devitt, J. Q. You, and F. Nori, Phys. Rev. A **97**, 022335 (2018).
[4] X. G. Wang, Z. Sun, and Z.D. Wang, *Phys. Rev. A* **79**, 012105 (2009).
[5] S. Arroyo-Camejo, A. Lazariev, S. W. Hell, and G. Balasubramanian, *Nature Commun.* **5**, 4870 (2014).
[6] Z. Han, Y. Dong, B. Liu, X. Yang, S. Song, L. Qiu, D. Li, J. Chu, W. Zheng, J. Xu, T. Huang, Z. Wang, X. Yu, X. Tan, D. Lan, M.H. Yung, and Y. Yu, *arXiv:2004.10364* (2020).
[7] Z.-Y. Xue, J. Zhou, and Z. D. Wang, *Phys. Rev. A* **92**, 022320 (2015).
[8] P. Z. Zhao, G. F. Xu, and D. M. Tong, *Phys. Rev. A* **99**, 052309 (2019).
[9] K. Nagata, K. Kuramitani, Y. Sekiguchi, and H. Kosaka, *Nat. Commun.* **9**, 1-10 (2018).
[10] B. B Zhou, P. C Jerger, V. O. Shkolnikov, F. J. Heremans, G. Burkard, and D. D. Awschalom, *Phys. Rev. Lett.* **119**, 140503 (2017).
[11] M. A. Kowarsky, L. C. L. Hollenberg, and A. M. Martin, *Phys. Rev. A* **90**, 042116 (2014).
[12] J. Lu and L. Zhou, *EPL* **102**, 30006 (2013).
[13] F. Jelezko and J. Wrachtrup, *J. Phys.: Condens. Matter* **18**: S807-S824 (2006).
[14] F. Dolde, I. Jakobi, B. Naydenov, N. Zhao, S. Pezzagna, C. Trautmann, J. Meijer, P. Neumann, F. Jelezko, and J. Wrachtrup, *Nature Phys.* **9**, 139 (2013).
[15] H. R. Wei and G. L. Long, *Phys. Rev. A* **91**, 032324 (2015).
[16] Y. Wu, Y. Wang, X. Qin, X. Rong, and J. Du, *npj Quantum Info.* **5**, 1–5 (2019).
[17] J. M. Rios. *Quantum manipulation of nitrogen-vacancy centers in diamond: from basic properties to applications*. Harvard University, 2010.
[18] D. A. Hopper, H. J. Shulevitz, and L. C. Bassett, *Micromachines* **9**, 437 (2018).
[19] V. A. Mousolou, *Phys. Rev. A* **96**, 012307 (2017).
[20] C. Zu, W.B. Wang, L. He, W.G. Zhang, C.Y. Dai, F. Wang, and L. M. Duan, *Nature* **514**, 72 (2014).
[21] M. Haruyama, S. Onoda, T. Higuchi, W. Kada, A. Chiba, Y. Hirano, T. Teraji, R. Igarashi, S. Kawai, H. Kawarada, *et al.*, *Nat. Commun.* **10**, 1–9 (2019).
[22] B.H. Huang, Y.-H. Kang, Z.C. Shi, J. Song, and Y. Xia, *Ann. Phys.* **530**, 1800179 (2018).
[23] T. H. Xing, X. Wu, and G. F. Xu, *Phys. Rev. A* **101**, 012306 (2020).
[24] P. Z. Zhao, K. Z. Li, G. F. Xu, and D. M. Tong, *Phys. Rev. A* **101**, 062306 (2020).
[25] D. Mahony and S. Bhattacharyya, *Appl. Phys. Lett.* **118**, 204004 (2021); F. Mazhandu, K. Mathieson, C. Coleman, and S. Bhattacharyya, *Appl. Phys. Lett.* **115**, 233501 (2019).
[26] K. Xu, W. Ning, X. J. Huang, P. R. Han, H. Li, Z. B. Yang, D. Zheng, H. Fan, and S. B. Zheng, *arXiv:2009.03610*.
[27] X. Liu, X. J. Liu, and J. Sinova, *Phys. Rev. B* **84**, 035318 (2011).
[28] M. W. Doherty, N. B. Manson, P. Delaney, and L. C. L. Hollenberg, *N. J. Phys.* **13**, 025019 (2011).
[29] P. Roushan *et al.*, Nat. Phys. **13**, 146 (2016); P. Roushan *et al.*, *Nature* **515**, 241 (2014).
[30] D. W. Wang, C. Song, W. Feng, H. Cai, D. Xu, H. Deng, D. Zheng, X. Zhu, H. Wang, S. Zhu, and M. O. Scully, *Nat. Phys.* **15**, 382 (2019).
[31] S. Bhattacharyya and S. Bhattacharyya, *J. Appl. Phys.* **130**, 034901 (2021); S. Bhattacharyya, D. Mtsuko, C. Allen, and C. Coleman, New J. Phys. **22**, 093039 (2020).
[32] T. Chen, P. Shen, and Z. Y. Xue, *Phys. Rev. Appl.* **14**, 034038 (2020).
[33] S. B. Zhang, W. B. Rui, A. Calzona, S. J. Choi, A. P. Schnyder, and B. Trauzettel, *Phys. Rev. Res.* **2**, 043025 (2020)
[34] G. F. Xu, P. Z. Zhao, Erik Sjöqvist, and D. M. Tong, *Phys. Rev. A* **103**, 052605 (2021).
[35] Y. Dong, S.C Zhang, Y. Zheng, H. B. Lin, L. K. Shan, X. D. Chen, W. Zhu, G. Z. Wang, G. C. Guo, and F. W. Sun, *arXiv:2102.09227v1*.
[36] S. Li, T. Chen, and Z. Y Xue, *Adv. Quantum Technol.* **3**, 1-7, 2000001 (2020).
[37] G. A. Yan and H. Lu, *Int. J. Theo. Phys.* **59**, 2223 (2020).
[38] X. Zhu, Y. Matsuzaki, R. Amsüss, K. Kakuyanagi, T. Shimo-Oka, N. Mizuochi, K. Nemoto, K. Semba, W. J. Munro, and S. Saito, *Nat. Commun.* **5**:3524 (2014).
[39] D. G. Lai, X. Wang, W. Qin, B. P. Hou, F. Nori, and J. Q. Liao, *Phys. Rev. A* **102**, 023707 (2020); D. G. Lai, J. F. Huang, X. L. Yin, B. P. Hou, W. Li, D. Vitali, F. Nori, and J. Q. Liao, *Phys. Rev. A* **102**, 011502(R) (2020).
[40] Y. H. Chen, W. Qin, R. Stassi, X. Wang, and F. Nori, arXiv:2012.06090v2 (2021); X. Gu, A.F. Kockum, A. Miranowicz, Y.X. Liu, F. Nori, *Phys. Rep.* **718-719**, pp. 1-102 (2017).
[41] D. Churochkin, R. McIntosh, and S. Bhattacharyya, *J. Appl. Phys.* **113**, 044305 (2013).

# Holonomic control of a three-qubits system in an NV center using a near-term quantum computer

Shaman Bhattacharyya and Somnath Bhattacharyya
Nano-Scale Transport Physics Laboratory, School of Physics, University of the Witwatersrand, Johannesburg, 1 Jan Smuts Avenue, Wits 2050, South Africa

**Methods:** All the gates that are used to perform the holonomic control can be expressed as single qubit U1, U2, and U3 gates. No multi-qubit gates have been used. This is because all the gates have to be consistent with the quantum Rabi model and also for a gate to be holonomic the gate must cause the qubit to rotate in any arbitrary axis while creating a geometric phase. Using a multi-qubit gate introduces the Dicke effect since the number of qubits is greater than 1. This is beyond the scope of our work. The fidelity of single-qubit gates is usually higher than a multi-qubit gate for these types of experiments.

**Two qubits:** After making the rotating approximation, the Hamiltonian acting on this system can be expressed in the rotating frame as

$$H = \sum_{i=1}^{4} |\omega_i|i><i| + \frac{i}{2}(\Omega_{P_1}e^{-iV_{P_1}t} - \Omega_{P_2}e^{-iV_{P_2}t})(|1><3|-|1><4|)$$
$$-\frac{i}{2}(\Omega_{S_1}e^{-iV_{S_1}t} - \Omega_{S_2}e^{-iV_{S_2}t})(|2><3|+|2><4|) \quad (S1)$$

The Hamiltonian in the interaction frame can be expressed as

$$H_I = \begin{pmatrix} 0 & 0 & i\Omega_{p1}/2 & i\Omega_{p2}/2 \\ 0 & 0 & -i\Omega_{s1}/2 & -i\Omega_{s2}/2 \\ -i\Omega_{p1}^*/2 & i\Omega_{s1}^*/2 & \Delta_1 & 0 \\ -i\Omega_{p2}^*/2 & i\Omega_{s2}^*/2 & 0 & -\Delta_2 \end{pmatrix} \quad (S2)$$

The norm of the return amplitudes for the 2 qubits (4 levels). NV center as functions of time as described by the 4 level interaction Hamiltonian H for the states |00>, |01>, |10>, and |11>, respectively. It is important to note that the amplitude of states |10> and |11> are the same [Fig. S1a) and b)].

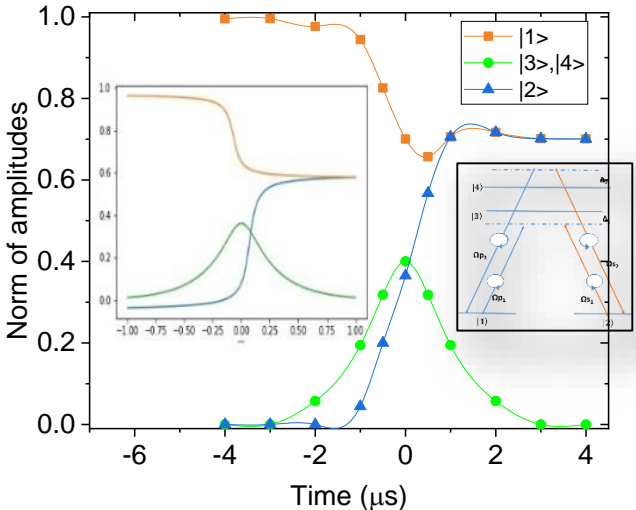

**Fig S1** The norm of amplitudes for state vs. time shows the evolution of the states over time illustrating a $\pi/2$ rotation of a four-level NV centre on a quantum computer. The blue line and triangles represent the state |2>. The orange line and squares represent the state |1>. The green line and circles represent the states |3> and |4>. The states |1> and |2> are both degenerate. The Rabi frequencies of the NV center are both shown as functions of time which are already shown in Ref. 9. **Inset:** The evolution of the states over time according to the model described by the interaction Hamiltonian given in Eqn. S1 and S2 and the energy level diagram [reproduced the results of Ref. 12].

To perform the calculations, we assume that the system is initially at the state |1>. Then we solve the 4 level interaction Hamiltonian using a time-dependent Schrödinger equation. We set ω = 20 MHz and α = 1.1 MHz. We then plot the norms of the amplitudes for all 4 states. It can be noted that the amplitudes for the |3> and |4> states follow the same trend and no populations of these two states exist both before and after the holonomic control pulses. For state |1> the amplitude decreases from 1 to about $1/\sqrt{2}$. The amplitude for state 01 increases to $1/\sqrt{2}$ from 0. This result is consistent as it shows that the qubit was rotated by $\pi/2$ about the x-axis. Fig. S1 and inset show the evolution of the system over time to the exact equal superposition required after a $\pi/2$ rotation. The link between the four-level model and the states depicted here can be shown on two Bloch spheres, with $|1\rangle = |00>, |2\rangle = |01>, |3\rangle = |10>,$ and $|4\rangle = |11>$.

From the plotted return probabilities of each of the states in Fig. 2b) (in the main text) for the period over which the pulses are applied, it can be seen how the holonomic rotation of the one qubit is related to the rotation of the other qubit when the paths are concatenated. The return probabilities of the |3> and |4> states (shown by the orange line in Fig. 2b) over this period start at zero, rise to a maximum at $t = 0$, and fall off to zero again, indicating a rotation about a full-loop with a solid angle of $\pi/2$. This geometric phase corresponds to the $\pi/2$ $R_x$ rotation of the first qubit, as seen by the return probabilities of the |1> and |2> states – the $P_r$ of the |1> state is initially at 1, but decreases to $1/\sqrt{2}$, while that of the |2> state is increased from 0 to $1/\sqrt{2}$ after the pulses have been applied. This agrees with the result from Lu *et al.* who demonstrated such a two-qubit holonomic control using a 4 level NV center system [12]. The simulated reproduction of these results resembles those of Lu *et al.* more closely than the emulated reproduction of the results, as the quantum emulator used does not have NV center qubits, but rather uses superconducting qubits [12]. This, along with other sources of error such as noise, is why the discrepancy in the emulated results is higher than in the simulated results. This discrepancy is further shown in Fig. S1 and inset, where the population of the |00> state is plotted with time after the concatenation of the rotation paths. In the absence of noise, there is very little fluctuation in the $P_r$ of the |1> state (the orange line), but as soon as noise is introduced in the emulation, the population of the |1> state oscillates due to the geometric phase with a magnitude that depends on the level of degeneracy. The comparison to the case with no noise is important as it allows the phase to be indirectly measured.

**Two qubit holonomic gates:**
Abelian phases arise from non-degenerate systems due to the adiabatic evolution of the phase Hamiltonian. The exact opposite is true for degenerate states. An experiment was conducted and it was discovered that the nature of such geometric phases is non-Abelian. As mentioned in the introduction NV centers are an ideal choice for holonomic control due to their coherence times at room temperatures. Another key advantage of NV centers is the fact that the sublevels of the NV center can be manipulated via an external magnetic field that allows for the variation in degeneracy between eigenstates. Now degeneracy is possible for SU(2) and higher-order systems only. In this simulation, the axis of the first qubit is located in the nitrogen atom while the axis of the second qubit extends from the nitrogen atom to the adjacent vacancy. We use the lab frame reference for the qubit states. Without an external magnetic field, the ground states ±1 states are degenerate. A magnetic field applied to the z-axis of the second qubit induces Zeeman splitting. If no magnetic field is applied the resulting phase is Abelian since the state $m_s = 0$ is non-degenerate. To demonstrate the phase creation in this work we start by letting the ground states of the first qubit (|1> and |2>, respectively) to be degenerate. Now in an experiment degeneracy is maintained by rotating the NV center and the magnet together. In IBM QE we use two qubits to represent the 4 levels of the NV center. Just like with the single-qubit scenario a unitary operator U is implemented keeping one of the rotation angles $\theta$ or $\varphi$ constant. The operator U is expressed in terms of $R_X$, $R_Y$, and $R_Z$ gates in IBM QE because these operations are unitary and are expressed in terms of both $\theta$ and $\varphi$. We create two different rotation paths. We implement each path on each qubit. The starting point of the paths is created by applying an $R_X$ ($\pi/3$) rotation which is implemented by setting $\theta$ to -$\pi/2$ and $\varphi$ to $\pi/3$ on both qubits. The first path is a complete azimuthal rotation around the Bloch sphere. The phase matrix is approximated as: $U=$ ($-0.609+0.321i$ $0.786+0.1i$ $-0786+$ $0.1i$ $-0.609+0.321i$). The second path is a complete loop defined by $\theta$ and $\varphi$. The closed-loop is created by a complete rotation around the y-axis of the second qubit. The coordinates of this rotation given in terms of $\varphi$ and $\theta$ are given by $(\theta, \varphi) = [(\pi/3, 0), (0, \pi/3), (-\pi/3, 0), (0, -\pi/3)]$. Now both paths accumulate a geometric phase. Since the excited states are non-degenerate and only d$\varphi$ or d$\theta$ is nonzero the resulting geometric phase is Abelian. We also interchange the paths to observe any difference in the geometric phase. Now a disadvantage of SU(2) and higher-order systems is that the resulting geometric phase cannot be directly observed. In IBM QE we attempt to show this geometric phase by measuring the qubits in the |z> basis to obtain the density of the ground states. We also determine the expected density of states in the absence of the geometric phase and decoherence.